\documentclass[twocolumn,pre,floatfix,showpacs]{revtex4}

\usepackage{epsfig}
\usepackage{amsmath}
\usepackage{times}

\def\be{\begin{equation}} 
\def\ee{\end{equation}} 
\def\ep{\epsilon}
\def\te{\tilde{\epsilon}}

\begin{document}

\title{Electrorotation in graded colloidal suspensions}
\author{J.\,P. Huang}
\affiliation{Department of Physics, The Chinese
University of Hong Kong, Shatin, NT, Hong Kong, and
Biophysics and Statistical Mechanics Group, Laboratory of Computational Engineering,
Helsinki University of Technology,  P. O. Box 9203, FIN-02015 HUT, Finland}

\author{K.\,W. Yu}
\affiliation{Department of Physics, The Chinese
University of Hong Kong, Shatin, NT, Hong Kong}

\author{G.\,Q. Gu}
\affiliation{Department of Physics, The Chinese
University of Hong Kong, Shatin, NT, Hong Kong, and 
College of Information Science and Technology, East
China Normal University, Shanghai 200 062, China}

\author{Mikko Karttunen} 
\affiliation{Biophysics and Statistical Mechanics Group, 
Laboratory of Computational Engineering,
Helsinki University of Technology,  P. O. Box 9203, FIN-02015 HUT, Finland}

\begin{abstract}
 Biological cells can be treated as composites of graded material inclusions.
 In addition to biomaterials, graded composites are important in  
  more traditional materials science. In this article,
 we investigate the electrorotation (ER) spectrum of a graded
 colloidal suspension in an attempt to discuss its dielectric
 properties. For that, we use the recently obtained differential effective 
 dipole approximation (DEDA) and generalize it for non-spherical particles.
 We find that variations in the  conductivity profile may make the
 characteristic frequency red-shifted and have also an effect on
 the rotation peak. On the other hand, variations in the dielectric profile may enhance
 the rotation peak, but do not have any significant effect on the characteristic
 frequency. In the end, we apply our theory to fit experimental data
 obtained for yeast cells and find good agreement.
\end{abstract}

%%%%%%%%%%%%
% 82.70.-y Disperse systems; complex fluids
% 77.22.Gm Dielectric loss and relaxation
% 61.20.Qg Structure of associated liquids: electrolytes, molten salts, etc.
% 77.84.Nh Liquids, emulsions, and suspensions; liquid crystals 
% 77.22.-d Dielectric properties of solids and liquids
% 77.22.Ej Polarization and depolarization
% 77.84.Lf Composite materials
% 42.79.Ry Gradient-index (GRIN) devices 
%%%%%%%%%%%%%%%

\pacs{82.70.-y,77.22.Gm,77.22.-d,77.84.Lf}
\maketitle

\section{Introduction}

AC electrokinetic phenomena, i.e., electrorotation, dielectrophoresis
and traveling wave dielectrophoresis, have gained an increasing 
amount of attention during the past decade. This is due to their wide
range of applications from cancer research to 
identifying and separating parasites, cell populations
and viruses~\cite{Dalton:2001a,Jun:2000a}, and even to 
design of nanomotors~\cite{hughes00a,hughes:2002b}. 
Despite the number of applications,
there is a need for a theory that treats the different aspects
of electrokinetic phenomena on an equal footing starting from 
the general underlying physical principles~\cite{Chan}. 

In this article we concentrate on electrorotation.
Electrorotation (ER) is a phenomenon in which
an interaction between a rotating ac electric field~\cite{Zim,l6,l7} 
and suspended dielectric particles leads to a rotational motion of the particles.
This phenomenon was first observed experimentally over
four decades ago~\cite{l60} and during the last two
decades, ER has been increasingly employed as a sensitive tool for non-invasive
studies of a broad variety of microparticles, ranging from 
living cells to spores, seeds as well as synthetic
materials~\cite{Zim,l2,l3,l4,l5,JPCM}.

In electrorotation, a rotating ac electric field
induces a dipole moment inside a polarizable particle which rotates at the
angular frequency of the external field. If the frequency of the field increases,
the period of the rotating field becomes comparable to
the time scale related to the formation of a dipole.
To minimize its energy, the dipole  tries to line up with the field
but at high frequencies the particle is no longer able to follow the field.  
This leads to 
a lag in its response and a torque is induced by the 
interaction of the out-of-phase part of the induced dipole
moment and the external field. 
The torque
and, in turn, the rotating speed are proportional to the imaginary part of the
dipole factor (also called Clausius-Mossotti factor). In addition, the
rotating speed of a particle is inversely proportional to the dynamic
viscosity,  since the particle is suspended in a viscous medium and experiences
a dissipative drag force.

Different models have been suggested and used for
analyzing  experimental results, see e.g. Refs.~\cite{Chan,Gimsa:2001a} and
references therein. 
The most commonly used ones are the so-called
shell models. This approach was  first used by Fricke already in 1925~\cite{Fricke:1925a}. 
A model for a biological cell consisting of a single spherical
dielectric shell was introduced by 
Arnold and Zimmermann in 1982 to investigate the
rotation of a mesophyll protoplast~\cite{Zim}. 
Four years later, a two-layer
model consisting of two spherical shells is put forth to discuss
the electrorotation of a single plant cell~\cite{Fuhr1}. 
Due to the inhomogeneous
compartmentalization of cells, three-shell models have been used to
investigate ER of liposomes~\cite{Chan}.

These inhomogeneities can be studied in the framework of 
graded materials, i.e., materials with spatial gradients in their structure.
Biomaterials such as cells and bones are typical 
examples as their composition varies through the object. 
Gradations can also be used to control and improve the strength
and other properties, for a recent review see Ref.~\cite{Suresh:2001a}.
In biological and medical applications identification and separation 
of graded objects in micron scale (and below) is of great importance.
In these situations gradation is generic due to inhomogeneous compartments
of cells, the cytoplasm often contains organelles or vacuoles. 
These inhomogeneities are reflected in their dielectric properties.
Since differences and changes in the dielectric properties are intimately 
related to physical and chemical properties of cells (as well as other objects),
electrorotation and other electrokinetic phenomena allow identification of
particles and detection of compositional changes.

As far as one decade ago, Freyria {\it et
al.} observed a graded cell response when they experimentally study
cell-implant interactions~\cite{bio1}. Recently, motivated by
positional information of protein molecules, Honda and Mochizuki
performed computer
simulations of cell pattern formation showing that graded cells can be
generated by cell behavior involving differences in intercellular
repulsion~\cite{Hon}. 
To discuss such graded
cells, the above-mentioned multi-shell models fail to apply. It is
thus necessary to develop a new theory to study the effective
properties of graded composite materials under externally applied
fields. In a recent work~\cite{PLA-2}, we succeeded in
deriving the dipole factor after putting forth a differential
effective dipole approximation (DEDA).  
In this work, we generalize DEDA to non-spherical particles and apply it 
to investigate the ER spectrum of graded particles in an attempt
to investigate their dielectric behavior.
Finally, to demonstrate the validity of our model, 
we fit our theory to experimental
data~\cite{exper} and find good agreement.

\section{Formalism}

We consider an inhomogeneous biological cell or a particle of radius $a$
with complex dielectric constant profile $\te_1(r)$, embedded in a
host fluid of dielectric constant $\te_2$. 
Here $\te=\ep+\sigma/(2\pi i f)$, where $\ep$
denotes real part of the dielectric constant, $\sigma$ stands for 
conductivity, $f$ is the frequency
of an external field, and $i=\sqrt{-1}$. The dipole factor reflects the
polarization of a particle in surrounding medium.

In a recent work~\cite{PLA-2}, we derived the  dipole factor
for spherical particles by introducing a differential
effective dipole approximation (DEDA). The idea of DEDA can be 
summarized as follows: Consider a shell model for an inhomogeneous particle.
In DEDA one adds new shells of infinitesimal thickness to the particle under
consideration. Each of these cells have distance dependent dielectric constant. 
As the thickness of the layer approaches zero, i.e., $\mathrm{d}r \rightarrow 0$,
then the correction to the dipole factor due to it is infinitesimal and
one can obtain a differential equation for it -- hence the name differential
effective dipole approximation.
Next we extend DEDA to the general case of graded spheroidal particles. Then,
we study the effect of different conductivity and dielectric profiles 
to the electrorotation spectrum, and finally compare the theory to experiments.

We first consider a homogeneous dielectric
spheroid in a uniform electric field. The spheroidal inclusion has a
dielectric constant $\te_1$ and it is embedded in a host medium of dielectric
constant $\te_2$. A spheroidal particle can be parametrized
by $x^2+y^2+(z/h)^2=r^2$ and the dielectric profile for the graded
particle is given by $\te (r)$, where $0<r \le a'$ and $h$ is the aspect
ratio and $a'$ is the length of the semi principle axis along $x-(y)$~axis.
It is worth noting that $h>1$ denotes a prolate spheroid,
whereas $h<1$ indicates an oblate one. 
We
assume that the longest axis of the spheroid is along the $z-$axis with the
depolarization factor $L_z$, satisfying a sum rule $L_z+2L_{xy}=1$
where $L_{x(y)}$ is the depolarization factor along $x(y)-$axis. For a
homogeneous spheroid, the dipole factor along $z-$axis $b_{z}$ is
given by
\begin{equation}
b_z =\frac{1}{3}\frac{\te_1-\te_2}{\te_2+L_z (\te_1-\te_2)},
\label{bz}
\end{equation}
which measures the degree of polarization for a particle in an
external field. Next, we add a confocal spheroidal shell of dielectric
constant $\te$ to the spheroidal particle to make a coated spheroid.
Then, the dipole factor of the coated spheroid is~\cite{Sheng:1980a,Gao-JPCM}
\begin{equation}
b_{1z}=\frac{1}{3}\frac{(\te-\te_2)+3x_1\rho [\te+L_z (\te_2-\te)]}
                 {[\te_2+L_z(\te-\te_2)]+3x_1\rho L_z(1-L_z)(\te-\te_2)},
\label{b1z}
\end{equation}
where $\rho$ is the volume ratio of the core with respect to the  whole spheroid, and
$$
%%\begin{equation}
x_1=\frac{1}{3}\frac{\te_1-\te}{\te+L_z(\te_1-\te)}.
%%\end{equation}
$$

This approach can be extended to more shells of different dielectric
constants, at the expense of obtaining more complicated expressions.
It is easy to check that $b_{1z}$ (Eq.~(\ref{b1z})) 
reduces to $b_z$ (Eq.~(\ref{bz})) when $\te = \te_1$.
Thus, the dipole factor remains unchanged if one adds a spheroidal
shell of the same dielectric constant.

Next, we consider an inhomogeneous spheroid with the dielectric
profile $\te (r)$. To establish DEDA, we mimic the graded profile
by a multi-shell construction by building up a dielectric profile by
gradually adding shells. We start with an infinitesimal spheroidal
core of dielectric constant $\te (0)$ and continue to add confocal
spheroidal shells of dielectric constant $\te (r)$ until $r=a'$ is
reached. At $r$, we have an inhomogeneous spheroid. To establish a
functional form, we replace the inhomogeneous spheroid by a
homogeneous spheroid of the same dipole factor, and the graded profile
is replaced by an effective dielectric constant $\bar{\ep} (r)$. Thus,
Eq.~(\ref{bz}) becomes 
\begin{equation}
b_z(r)=\frac{1}{3}\frac{\bar{\ep}(r)-\te_2}
                       {\te_2+L_z (\bar{\ep}(r)-\te_2)}.
\end{equation}
Next, we add a confocal spheroidal shell of infinitesimal thickness
$\mathrm{d}r$, and dielectric constant $\te (r)$. The dipole factor will
change according to Eq.(\ref{b1z}). The effective dielectric constant
$\bar{\ep}(r)$, being related to $b_z(r)$, will also change. To see
this, let us write $b_{1z}=b_z+{\rm d}b_z$ and take the limit ${\rm d}r\to 0$. 
We obtain a differential equation:
\begin{eqnarray}
\frac{{\rm d}b_z}{{\rm d}r} &=& -\frac{1}
{r\te_2\te (r)} \left[\te_2 (1+3b_z L_{1z})-(1-3L_z b_z)\te(r) \right]\nonumber\\
                            & &\left[\te_2 L_z(1+3b_zL_{1z})+L_{1z}(1-3b_zL_z)\te (r) \right],
\end{eqnarray}
where $L_{1z} = 1-L_z$.
The dipole factor of a graded spheroidal particle can be calculated by
solving the above differential equation with a given gradation profile
$\te (r)$. This nonlinear first-order differential equation can be
integrated, at least numerically, if we are given a profile for $\te
(r)$ and an initial condition for $b_z(r=0)$. It is worth noting that the
dipole factor is a tensorial quantity and the dipole factor along
$x(y)-$ axis $b_{x(y)}$ can be obtained by changing $L_z$ to $L_{x(y)}$ in
the equation above. It is important to note that substituting $L_z=1/3$ into
this equation yields the DEDA equation for spherical graded 
particles~\cite{PLA-2}, 
\begin{eqnarray}
\frac{{\rm d}b}{{\rm d}r} & = &  -\frac{1}{3r\te_2\te_1(r)}
              \left[(1+2b)\te_2-(1-b)\te_1(r)\right] \label{DEDA} \\ 
\, & & \left[(1+2b)\te_2+2(1-b)\te_1(r) \right], \nonumber
\end{eqnarray}
where $0<r\le a$, and $a$ is the radius of the spherical particle.
This is the expression we will use in our studies of the effect of conductivity and
dielectric profiles on the ER spectrum.

The dipole factor of a graded spherical particle can be
calculated by solving the above nonlinear
first-order differential equation (Eq.~(\ref{DEDA})) with a given profile $\te_1(r)$ and
the initial condition 
\be
b(r=0)=\frac{\te_1(0)-\te_2}{\te_1(0)+2\te_2}.   
\ee

If we apply a rotating field to the system, a torque between induced
dipole moment inside the cell and the electric field vector will set the
cell in a rotational motion. It is well known that the frequency-dependent rotation
speed $\Omega(f)$ is proportional to the imaginary part of the dipole
factor $-\mathrm{Im}[b(r=a)]$. The rotation speed is given by 
\be \Omega(f)=-\frac{\ep_2E_0^2}{2\eta}\mathrm{Im}[b(r=a)], 
\ee 
where $E_0$ is the magnitude of the rotating field and $\eta$ the viscosity
of the host fluid.  Hence, when $b(r=a)$ is solved by integrating
Eq.~(\ref{DEDA}), we can investigate the electrorotation spectrum.

In order to perform numerical calculations, we take the conductivity
and dielectric profiles to be
\begin{eqnarray}
\sigma_1(r)&=&\sigma_1(0) (r/a)^n,  \mathrm{\, \, \, \, \, \, \,} r \le a\nonumber\\
\ep_1(r)&=&\ep_1(0)+c(r/a),         \mathrm{\, \,}                r \le a\nonumber
\end{eqnarray}
where $n$ and $c$ are profile dependent constants. The profile is
quite physical in the sense that the conductivity can change rapidly
near the boundary of cell and a power-law profile prevails.  
The profile constant $n$ can take any positive
values and it can be larger than unity. From the above equations it is
clear that the conductivity is
largest at the boundary. This is the case in real systems as well, 
since the outermost shell
(e.g. cell wall) of a cell is always conductive and often has the
largest conductivity.  On the other hand, the dielectric constant may
vary only slightly and thus a linear profile suffices.  In particular, 
the dielectric constant at the center, i.e., $\ep_1(0)$, may
be larger than that at the boundary. Thus, we will choose $c \le 0$.
By integrating the dielectric profile, we obtain an average dielectric
constant $\ep_{av}$ for different values of $c$ by using a volume
average: 
\be \ep_1{}^{av}=\frac{\int_0^a\ep_1(r)r^2{\rm
d}r}{\int_0^ar^2{\rm d}r}.   
\label{eave}
\ee 
For the present dielectric profile,
we obtain $\ep_1{}^{av}=\ep_1(0)+3c/4$.

\section{Numerical results}

We are now in a position to perform numerical calculations. We set
$\sigma_1(0)=2.8\times 10^{-2}$\,S/m, $\ep_1(0)=75\ep_0$, $\ep_2=80\ep_0$,
$\sigma_2=2.8\times 10^{-4}$\, S/m for all of the computations. 
Parameter $\ep_0$ denotes the dielectric constant of
vacuum. Here, the numerical results and the comparison to experimental data is
limited to spherical cells only since we were not able to find 
appropriate experimental data describing non-spherical cells.
For the numerical computations, we used the fourth-order Runge-Kutta
algorithm with step size $h=0.01$ to solve the above nonlinear first-order
differential equation (Eq.~(\ref{DEDA})). It was verified that this step size
guarantees accurate numerics.
\begin{figure}[h]
\hspace*{-0.4cm}
\epsfig{file=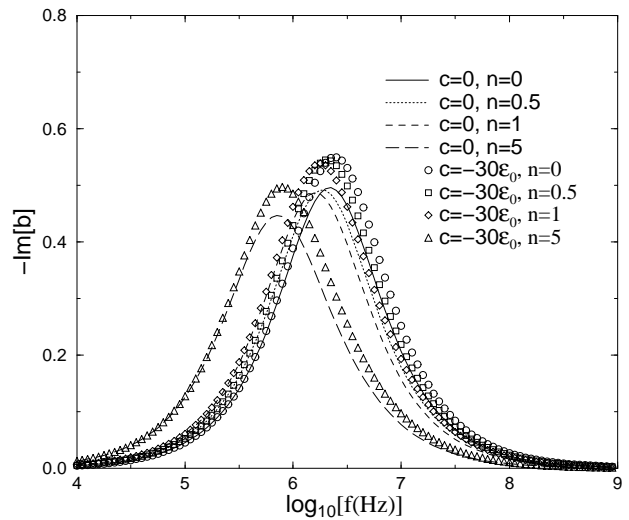,width=3.8in}
\caption{ER spectrum for profile constants $n$ at $c=0$ and $c=-30\ep_0$, respectively. 
The spectrum is given as the imaginary part of the dipole factor.}
\label{fig1}
\end{figure}

\begin{figure}[h]
\hspace*{-0.4cm}
\epsfig{file=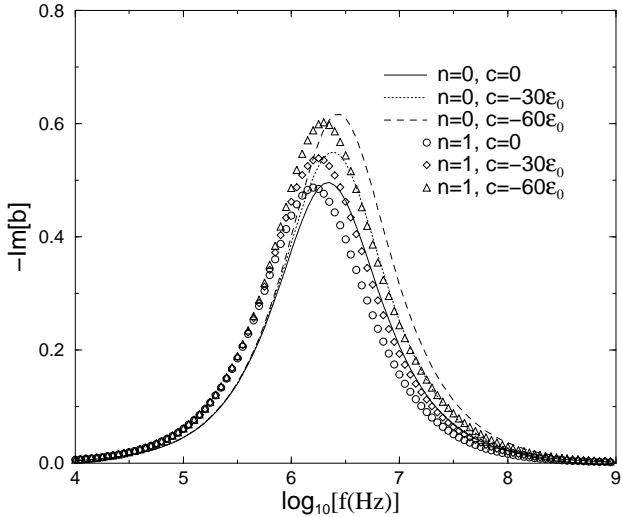,width=3.8in}
\caption{Same as in Fig.1, but for various $c$ at $n=0$ and $n=1$, respectively. }
\label{fig2}
\end{figure}

\begin{figure}[h]
\epsfig{file=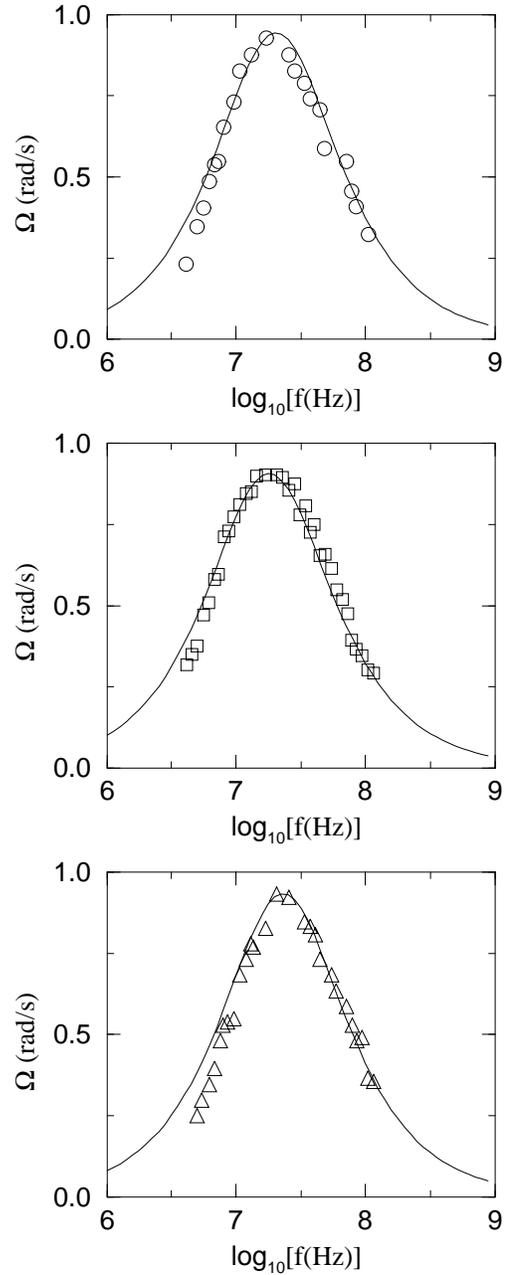,width=6.5cm}
\caption{Fitting of experimental ER spectrum of a yeast cell.
The data is extracted from Ref.~\cite{exper}. 
The symbols denote experimental 
data which is measured in water at the beginning, 
after 20 min, and after 40 min, respectively. The corresponding 
line denotes the results using the present model.}
\label{fig3}
\end{figure}

In Fig.~\ref{fig1}, the ER spectrum is investigated for various indices $n$ with
profile parameters
$c=0$ and  $c=-30\ep_0$. This corresponds to $\ep_1{}^{av}=75\ep_0$ and $52.5\ep_0$
calculated using Eq.~(\ref{eave}),
respectively. 
Figure~\ref{fig1} shows that the rotation peak is always present.
It is important to note that the characteristic frequency at which the rotation
speed reaches its maximum may be red-shifted, especially at large $n$ (e.g.
$n=5$) in  comparison with small $n$. Thus, the conclusion is that large
fluctuations in conductivity of graded particles may lead to a red-shift
of  the characteristic frequency.

In Fig.~\ref{fig2}, ER spectrum is investigated for $n=0$ and  $n=1$, 
using various slopes $c$. We use
$c=0$, $c=-30\ep_0$ and $c=-60\ep_0$ corresponding to
$\ep_1{}^{av}=75\ep_0$, $52.5\ep_0$ and $30\ep_0$, respectively. 
From the figure, it is evident that large spatial fluctuations in the 
dielectric constant may enhance the rotation peak.

From Figs.~\ref{fig1}~and~~\ref{fig2}, we find that spatial conductivity fluctuations 
are able to make the
characteristic frequency red-shifted and have a clear effect on the rotation peak.
In contrast, spatial dielectric constant fluctuations
may enhance the peak value, but their effect on the characteristic frequency
is very small. To summarize, we can theoretically obtain a typical ER
spectrum with lower characteristic frequency and high rotation speed by
adjusting conductivity and dielectric profiles.
This corresponds to the physical/experimental situation in the case of graded particles.

In  Figs.~\ref{fig1}~and~~\ref{fig2}, the particles under consideration are 
spherical, i.e.,  $L_z=1/3$. We should remark here that for non-spherical
particles, e.g. prolate $0<L_z<1/3$ and oblate
spheroids $1/3<L_z<1$, the same conclusions
hold (no figures shown here). 
In addition, it is also found that for
prolate (oblate) graded particles, smaller (larger) depolarization factor
($L_z$) can make the characteristic frequency significantly red-shifted.

Finally, to demonstrate our theory, we  fit our theory to experimental
data in Fig.~\ref{fig3}. The
data for yeast cells is extracted from the experiments of 
Reichle {\it et al.}~\cite{exper}. 
In Fig.~\ref{fig3}, from upper panel to lower panel, the
symbols denote experimental data of the ER spectrum of a yeast cell in water
measured at the beginning of the experiment, after 20 minutes, and after 40 minutes,
respectively. The corresponding continuous lines are the fitted result
using the present theory. As discussed earlier, 
the assumption of homogeneous compartments is not generally
justified  for biological
cells, and the cytoplasm often contains organelles or vacuoles. 
For this reason, the experimental data is typically fitted using 
a (two-)shell model. 
To improve fitting using cell models, one would have to  
adjust six parameters, i.e., real dielectric constant and conductivity of
two shells, as well as cell interior. In contrast, using on our theory, 
only two parameters need to be adjusted,
namely the profile parameters
$c$ and $n$.

For all of the three fittings, we set $\ep_2=80\ep_0$, $\sigma_2=1.1\times
10^{-3}$~S/m (from the experiment), $\ep_1(0)=75\ep_0$,
$\sigma_1(0)=0.28$~S/m, $E_0=10$~kV/m (from experiments),
$\eta=2.082\times 10^{-2}$~kg/(ms), and $a=5\,\mu$m (given by experiment).
For the upper panel, $c=-30\ep_0$ (namely $\ep_1{}^{av}=52.5\ep_0$) and
$n=0.5$. For the middle panel, $c=-20\ep_0$ ($\ep_1{}^{av}=60\ep_0$)
and $n=0.8$. For the lower panel, $c=-25\ep_0$ 
($\ep_1{}^{av}=56.25\ep_0$) and $n=0.05$.

As seen in Fig.~\ref{fig3}, our theory performs very well. 
Due to
the inhomogeneous compartments of a yeast cell, we believe that the graded cell
model is a very choice.

\section{Discussion and conclusions}

In this article, we have generalized the differential effective dipole approximation
(DEDA) to non-spherical particles. This allows studies of shape effects, and indeed 
colloidal suspensions and many kinds of cells have non-spherical
shapes, e.g. red blood cells are oblate spheroids. 
In particular, for oblate spheroids larger depolarization factor leads to a 
red-shift of the characteristic frequency, for prolate spheroids the systems behaves
the opposite way~\cite{JPCM}.
To check the validity of DEDA, it was compared against exact solutions which can be 
obtained for a power-law profile~\cite{gunpublished} and a linear profile~\cite{dongunpublished}
by solving the Laplace equation for the local electric field.
In both cases DEDA produced the same results thus showing that it is essentially exact.

In this article, we have applied DEDA to discuss the electrorotation 
spectrum of graded colloidal suspension in an attempt to discuss its dielectric behavior. 
As a result, we find that changes in the 
conductivity profile can affect both the characteristic frequency (red-shift)
and rotation peak,
whereas changes in the dielectric profile may enhance the rotation peak only. 
Consequently, we may
conclude that the graded behavior does play an important role in the
dielectric properties of graded suspensions. We used our theory (for spherical cells) 
to fit results obtained by Reichle {\it et al.}~\cite{exper} for yeast cells
and obtained excellent agreement. It would be interesting to compare this theory
to experimental results for non-spherical cells as well, and hopefully new experiments
will be performed in that direction.

Our work may be extended to discuss the ER spectrum of a pair of particles
by taking into account the multiple image effect~\cite{PRE2,jphuang02a} and to
investigate high concentration suspensions by discussing the local-field
effects~\cite{PLA,Gao}. A generalization which includes both 
of these effects is of particular interest.
In addition to ER spectrum, our theory may be used to discuss
other electrokinetic effects, e.g
dielectrophoresis~\cite{1978,jphuang02a} and electro-orientation~\cite{1966}.

\acknowledgments
This work has been supported by the Research Grants Council of the Hong
Kong SAR Government under project number CUHK 4245/01P, and by the
Academy of Finland Grant No.~54113 (M.\,K.) and the Finnish Academy of Science
and Letters (M.\,K.).

%%\bibliographystyle{apsrev}
%%\bibliography{dep}

\end{document}